\documentclass[preprint2]{aastex6}
\usepackage{amsmath,multirow}
\usepackage{natbib}
\usepackage{longtable}

\newcommand{\bootes}{Bo\"{o}tes}
\newcommand{\xbootes}{XBo\"{o}tes}
\newcommand{\xcos}{{\it XMM}-COSMOS}
\newcommand{\swift}{XRT-SDSS}

\newcommand{\lx}{L_{\rm X}}

\newcommand{\lsix}{L_{6\micron}}
\newcommand{\lxlsix}{$L_{\rm X}-L_{6\micron}$}
\newcommand{\lxlmir}{$L_{\rm X}-L_{\rm MIR}$}

\newcommand{\ergs}{$\textmd{erg}\;\textmd{s}^{-1}$}

\newcommand{\aox}{$\alpha_{\rm OX}$}

\newcommand{\fluxcgs}{ergs~s$^{-1}$~cm$^{-2}$}
\newcommand{\xmm}{\hbox{\textit{XMM}-Newton}}
\newcommand{\swiftbat}{\hbox{\textit{Swift}/BAT}}

\shorttitle{$L_{\rm X}$-$L_{\rm MIR}$ relation for type 1 quasars}
\shortauthors{Chen et al.}
\slugcomment{ApJ accepted}

\begin{document}

\title{The X-ray and Mid-Infrared luminosities in Luminous Type 1 Quasars}
\author{Chien-Ting J. Chen\altaffilmark{1,2}}
\author{Ryan C. Hickox\altaffilmark{2}}
\author{Andrew D. Goulding\altaffilmark{3}}
\author{Daniel Stern\altaffilmark{4}}
\author{Roberto Assef\altaffilmark{5}}
\author{Christopher S. Kochanek\altaffilmark{6}}
\author{Michael J.~I. Brown\altaffilmark{7}}
\author{Chris M. Harrison\altaffilmark{8}}
\author{Kevin N. Hainline\altaffilmark{2,9}}
\author{Stacey Alberts\altaffilmark{9}}
\author{David M. Alexander\altaffilmark{8}}
\author{Mark Brodwin\altaffilmark{10}}
\author{Agnese Del Moro\altaffilmark{11}}
\author{William R. Forman\altaffilmark{12}}
\author{Varoujan Gorjian\altaffilmark{4}}
\author{Christine Jones\altaffilmark{12}}
\author{Stephen S. Murray\altaffilmark{12}}
\author{Alexandra Pope\altaffilmark{13}}
\author{Emmanouel Rovilos\altaffilmark{8}}

\altaffiltext{1}{Department of Astronomy and Astrophysics, Pennsylvania State University, University Park, PA 16802, USA; ctchen@psu.edu}
\altaffiltext{2}{Department of Physics and Astronomy, Dartmouth College, 6127 Wilder Laboratory, Hanover, NH 03755, USA;}
\altaffiltext{3}{Princeton University, Department of Astrophysical
  Sciences, Ivy Lane, Princeton, NJ 08544, USA}
\altaffiltext{4}{Jet Propulsion Laboratory, California Institute of
  Technology, 4800 Oak Grove Dr., Pasadena, CA 91109, USA}
\altaffiltext{5}{N\'ucleo de Astronom\'ia de la Facultad de
  Ingenier\'ia, Universidad Diego Portales, Av. Ej\'ercito Libertador
  441, Santiago, Chile}
\altaffiltext{6}{Department of Astronomy, Ohio State University, 140 West 18th Avenue, Columbus, OH 43210}
\altaffiltext{7}{School of Physics, Monash University, Clayton 3800, Victoria, Australia.}
\altaffiltext{8}{Centre for Extragalactic Astronomy, Department of Physics, Durham University, South Road,
  Durham, DH1 3LE, United Kingdom}
\altaffiltext{9}{Steward Observatory, University of Arizona, Tucson,
  AZ 85721, USA}
\altaffiltext{10}{University of Missouri, 5110 Rockhill Road, Kansas City, MO 64110, USA}
\altaffiltext{11}{Max-Planck-Institut f\"{u}r Extraterrestrische Physik (MPE), Postfach 1312, D85741, Garching, Germany}
\altaffiltext{12}{Harvard-Smithsonian Center for Astrophysics, 60
  Garden Street, Cambridge, MA 02138.}
\altaffiltext{13}{Department of Astronomy, Amherst, University of
  Massachusetts, Amherst, MA 01003, USA}

\begin{abstract}
Several recent studies have reported different intrinsic correlations between
the AGN mid-IR luminosity ($L_{\rm MIR}$) and the rest-frame $2-10$ keV
luminosity ($L_\mathrm{X}$) for luminous quasars.
To understand the origin of the difference in the observed $L_\mathrm{X}-L_{\rm MIR}$ relations,
we study a sample of 3,247 spectroscopically confirmed type 1 AGNs collected
from Bo\"{o}tes, {\it XMM}-COSMOS, {\it XMM}-XXL-North, and the SDSS quasars in
the {\it Swift}/XRT footprint spanning over four orders of magnitude in
luminosity. We carefully examine how different observational constraints impact
the observed $L_\mathrm{X}-L_{\rm MIR}$ relations, including the inclusion of
X-ray non-detected objects, possible X-ray absorption in type 1 AGNs, X-ray flux
limits, and star formation contamination. We find that the primary factor
driving the different \lxlmir\ relations reported in the literature is the X-ray
flux limits for different studies. When taking these effects into account, we
find that the X-ray luminosity and mid-IR luminosity (measured at rest-frame
$6\micron$, or $L_{6\micron}$) of our sample of type 1 AGNs follow a bilinear
relation in the log-log plane:
$\log L_X =(0.84\pm0.03)\times\log L_{6\micron}/10^{45}\mathrm{erg\;s^{-1}} +
(44.60\pm0.01)$ for  $L_{6\micron} < 10^{44.79}\mathrm{erg\;s^{-1}} $,
and $\log L_X = (0.40\pm0.03)\times\log L_{6\micron}/10^{45}\mathrm{erg\;s^{-1}}
+(44.51\pm0.01)$ for $L_{6\micron} \geq 10^{44.79}\mathrm{erg\;s^{-1}} $.
This suggests that the luminous type 1 quasars have a shallower $L_\mathrm{X}-L_{\rm MIR}$
correlation than the approximately linear relations found in local Seyfert galaxies.
This result is consistent with previous studies reporting a luminosity-dependent
\lxlmir\ relation, and implies that assuming a linear $L_\mathrm{X}-L_{\rm MIR}$ relation to infer
the neutral gas column density for X-ray absorption might overestimate the column densities in luminous quasars.
\end{abstract}

\maketitle

\section{Introduction}\label{sec:intro}
X-ray and mid-IR emission are both excellent tracers of supermassive black hole
(SMBH) accretion activities. Since active galactic nucleus (AGN) emission at
these wavelengths is less susceptible to the presence of obscuring material
compared to optical wavelengths \citep[e.g.][]{corr16}, studying the correlation
between the X-ray and mid-IR luminosities of AGNs is crucial for understanding
the dust-enshrouded phase of galaxy-SMBH coevolution
\citep[e.g.][]{dima05qso,hopk06apjs,gill07cxb,some08bhev,trei09}.
There are now a range of studies examining the correlation between AGN X-ray
and mid-IR luminosities. Some of these works found that the mid-IR ($L_{\rm MIR}$)
and X-ray luminosities ($L_{\rm X}$) follow an almost linear relation in low
redshift, low luminosity AGNs
\citep[e.g.][]{lutz04irx,alex08compthick,gand09seyfir,luss11,asmu15}.
However, it is  not clear whether such a linear relation holds for more luminous
AGNs (i.e., quasars\footnote{We refer to AGNs with bolometric luminosity
($L_\mathrm{bol}$) more luminous than $10^{45}$ erg s$^{-1}$ as ``quasars''.}).
Notably, studies comparing $L_{\rm X}$ to the UV luminosity for AGNs found out
that the ratio between the X-ray and UV luminosities rapidly decreases with
increasing UV luminosity for type 1 AGNs \citep[e.g.][]{tana79,stra05,luss10}.
Since the rest-frame mid-IR emission of AGNs originates from the hot dust heated
by the UV photons from the SMBH accretion disk, understanding the $L_{\rm X}-L_{\rm MIR}$
relation for luminous AGNs is also crucial for understanding the structure of
the hot dust surrounding the central SMBH as well as the AGN accretion physics.

The local, linear \lxlmir\ relation is illustrated by the results of
\cite{gand09seyfir}, who found that the spatially resolved nuclear $L_{\rm MIR}$
and $L_{\rm X}$ for local Seyfert galaxies are almost linearly correlated.
Recently, \cite{asmu15} have extended this work to a number of more luminous AGNs
from the \hbox{9-month} \swiftbat\ catalog \citep{tuel08bat} and archival local
AGNs with high spatial resolution mid-IR observations \citep{asmu14}. \cite{asmu15}
found that the luminous AGNs in their sample have slightly more X-ray emission
than the value predicted by the local linear relation between $L_{\rm MIR}$ and
$L_{\rm X}$. While their result was only suggestive due to the limited size of
their sample, it is supported by the study of higher redshift AGNs selected
from the Bright Ultra-hard {\it XMM}–Newton survey \citep{mate15}.
However, some studies have also reported a luminosity dependent \lxlmir\
relation for luminous quasars, including the study of high redshift AGNs in
COSMOS by \cite{fior09obsc} and the compilation of SDSS DR5 AGNs spanning a wide
luminosity range studied by \cite{ster15}.

The lack of consensus on the universality of the \lxlmir\ correlations might
be due to various observational limitations. In particular, for
surveys such as COSMOS, the limited survey volumes restrict
the number of rare AGN detected at the highest luminosities.
On the other hand, wide-area surveys have shallower flux limits,
making them less likely to detect fainter sources and higher redshift
sources.  Thus, the \lxlmir\ correlations could also be biased if
the X-ray non-detected objects are not taken into account.

To understand whether such biases might affect the observed \lxlmir\ relations,
we compile four different type 1 AGN samples spanning a wide range of survey areas
and X-ray flux limits to investigate the intrinsic relationship between AGN mid-IR
and X-ray emission over a wide dynamic range in luminosity.

To minimize the contamination from star formation related processes and the
stellar emission in the host galaxy, we focus on luminous objects that are
spectroscopically confirmed as type 1 AGNs. We use type 1 AGN samples from
the AGN and Galaxy Evolution Survey \citep[AGES,][]{AGEScatalog} in the \bootes\
survey region, the publicly available AGN samples from the {\it XMM}-COSMOS survey \citep{luss10},
the {\it XMM}-XXL North survey \citep{pier16xxl,menz16xxl,liu16xxl}, and
the SDSS DR5 quasars with serendipitous {\it Swift}/XRT observations \citep{wu12dr5}.

This paper is organized as follows. In \S\ref{sec:lxlmir_sample} we describe the
multi-wavelength data and the properties of each quasar catalog. In
\S\ref{sec:lums}, we discuss the derivations
of X-ray and mid-IR luminosities.
In \S\ref{sec:xdet} and \S\ref{sec:bias}, we discuss the \lxlsix\ correlation
and the possible biases that might affect the observed relations.
A discussion and a summary are given in \S\ref{sec:con}.
Throughout the paper, we use the Vega magnitude system and assume a
$\Lambda$CDM cosmology with $\Omega_m=0.3$, $\Omega_\Lambda=0.7$ and
$H_0=70$  km s$^{-1}$ Mpc$^{-1}$.

\section{The Type 1 Quasar Samples}\label{sec:lxlmir_sample}
To investigate the correlation between
X-ray luminosities and the mid-IR luminosities for type 1 AGNs with broad
optical emission lines, we focus on extragalactic survey regions with X-ray
observations and mid-IR observations from {\it Spitzer} or {\it WISE}. We select
four different samples with a wide range of survey area and flux limits in order
to understand the biases that might affect the observed \lxlmir\ relation.

\subsection{\bootes\ type 1 quasar sample}\label{subsec:bootes}

One primary source of quasars for this study is the \bootes\ multiwavelength survey,
which has a wide area ($9$ deg$^2$) and excellent multiwavelength coverage.
For this work, we use the 1,443 AGNs in the AGES catalog that are classified
as ``type 1'' based on spectroscopic observations from the Hectospec instrument
on the MMT observatory \citep[i.e., sources that are best-fitted by the SDSS quasar template, see][for details]{AGEScatalog}.
\par

To ensure that the AGNs studied in this work have minimal impact from the radio-loud
quasars that could have X-ray emission enhanced by the presence of relativistic
jets \citep[e.g.,][]{zamo81,wilk87,capp96,brin00}, we also use the Westerbork
Synthesis Radio Telescope (WSRT) observations of the \bootes\ region to eliminate
powerful radio AGNs. \cite{cat_wsrt} surveyed the central $\approx 7$ deg$^2$ of
the NDWFS field at $1.4$ GHz to a limiting flux of $\approx 0.1$ mJy and beam
size $13\arcsec\times27\arcsec$. For the 46 matches (within 2$\arcsec$) between
the WSRT radio sources and the AGES AGN catalog, we calculate their ``radio-loudness''
using a radio-loud definition of $R\geq 10$ \citep{kell89}. Radio-loudness is
defined as the ratio between the 5 GHz and optical {\it B}-band (rest-frame)
monochromatic luminosities, $R = L_{5\mathrm{GHz}}/L_{B}$. $L_{5\mathrm{GHz}}$
is derived from the WSRT observations at 1.4 GHz assuming a typical power-law
spectrum, $S_\nu \propto \nu^{-0.7}$. The rest-frame $L_{B}$ is derived using
the SED-fitting results described in \S\ref{subsec:lxl6um_lir}. Of the 46 WSRT
detected sources, 33 of them are radio-loud AGNs \citep[$R>10$,][]{kell89}. We
exclude the 33 radio-loud sources and focus on the remaining 1,410 radio-quiet
AGNs in the following analysis.

\bootes\ is also covered by the \xbootes\ survey, a 9.3 deg$^2$ mosaic of 126
short (5ks) {\it Chandra} ACIS-I images \citep{murr05,kent05} covering the entire
AGES field. XBo\"{o}tes contains 2,724 X-ray point sources with at least four
counts in the AGES survey region. Of those, 790 X-ray point sources are far from
bright stars and matched within $3.5''$ to the 1,410 type 1 AGNs with good
spectroscopic redshifts from AGES at $0.14<z<3.61$ \citep{kent05,hick09}, yielding an
X-ray detection fraction of $\sim 56\%$ for the type 1 AGNs. These X-ray point
sources have 0.5-7 keV luminosities of $10^{42} < L_X < 10^{45}$ erg s$^{-1}$
which are characteristic of moderate to luminous AGNs. \par

We also make use of the optical to near-IR broad-band photometry available in the
\bootes\ field, which includes optical photometry from the NOAO Deep Wide Field
Survey \citep[{\it Bw, R, I},][]{NDWFS}, near-IR NEWFIRM \citep[{\it J, H, Ks},][]{gonz10},
mid-IR SDWFS \citep[{\it Spitzer} IRAC,][]{ashb09sdwfs} and mid-IR observation
at $24\;\micron$ from {\it Spitzer} MIPS \citep{inst_mips}. An extensive description of the
multiband photometry extraction can be found in \cite{brow07red} and \cite{chun14}.\par

Another advantage of the \bootes\ survey region is that it is covered by the
Herschel Multi-tiered Extragalactic Survey \citep[HerMES,][]{hermes}.
The inclusion of the far-IR photometry makes it possible to constrain
the star formation rate even for luminous quasars
\citep[e.g.,][]{netz07qsosf,kirk12,mull12agnsf,chen15qsosf}, thus allowing for
more accurate measurements of mid-IR AGN luminosities that could be contaminated
by star formation processes. For this work, we adopt the SPIRE 250$\micron$
photometry from \cite{albe13stack}.
For the 1,410 type 1 AGNs in AGES, $\approx 15\%$ of them are detected
by SPIRE at $250\micron$. For these far-IR detected AGNs, we carefully examine
the resulting SED fits in \S\ref{subsec:lxl6um_lir} and their \lxlmir\ relation
in \S\ref{subsec:lxmir_host}.

\begin{deluxetable*}{lcccccc}
\tabletypesize{\scriptsize}
\tablecolumns{6}
\tablewidth{0pt}
\tablecaption{Survey Properties}\label{tbl:survey}
\tablehead{
\colhead{} &
\colhead{\bootes\ } &
\colhead{\xcos\ } &
\colhead{\swift\ (10 ks)} &
\colhead{\swift\ (5 ks)} &
\colhead{XXL-North}}
\startdata
X-ray survey area             & 9.3 deg$^2$ & 2. deg$^2$ & \nodata$^a$ & \nodata  & 25 deg$^2$\\
(1) \# of type 1 AGN              & 1410        & 322        & 241         & 362      & 1153 \\
(2) {\it z}                       & 0.14-4.58   & 0.10-4.25  & 0.08-3.68   & \nodata  & 0.06-5.0\\
(3) $\left<m_r\right>$            & 21.0        & 21.5       & 19.0        &  19.0    & 20.7 \\
(4) X-ray energy range            & 0.5-7 ({\it Chandra}) & 0.5-10 ({\it XMM-Newton}) & 0.3-10 ({\it Swift}/XRT) & \nodata & 0.5-10 ({\it XMM-Newton}) \\
(5) 0.5--7 keV X-ray flux limit   & 7.8         & 1.0        & 25.0        & 50.0     & 11.1 \\
(6) X-ray detection fraction      & $56\% $ & $100\% $ & $83 \%$ & $70\%$  & $100\%$\\
(7) Median $L_{\rm X} (2-10 {\rm keV})$ & $44.23$ & $44.15$ & $44.66$ & $44.66$ & $44.26$\\
\enddata
\tablecomments{
$(1)$ Number of AGNs in each sample; see \S\ref{sec:lxlmir_sample}.\\
$(2)$ Redshift range.\\
$(3)$ Median {\it i}-band magnitude of each sample. \\
$(4)$ Observed frame X-ray energy range in keV.\\
$(5)$ X-ray flux limit at 0.5--7 keV in 10$^{-15}$ erg s$^{-1}$ cm$^{-2}$.
The flux limits of \xcos\ and the \swift\ catalogs have been converted to 0.5-7 keV
assuming a $\Gamma=1.8$ power-law SED and Galactic extinction.\\
$(6)$ X-ray detection fraction.\\
$(7)$ Median $L_{\rm X} (2-10 {\rm keV})$ for X-ray detected sources in $\log$ erg s$^{-1}$.
$^a$: The true survey area of the \swift\ catalog is not well-constrained due to
the varying X-ray exposure time of the catalog. \\
$^b$: The magnitude limit for target selection of XXL-N is based on {\it r}-band photometry.\\
}
\end{deluxetable*}

\subsection{XMM-COSMOS X-ray AGN sample}\label{subsec:cosmos}
Since \xbootes\ is a relatively shallow X-ray survey, we supplement it with the publicly available \xcos\ catalog of X-ray selected type 1 AGNs from \cite{luss10}. The \cite{luss10} catalog contains 545 X-ray AGNs of which 322 have secure spectroscopic redshift measurements and broad emission line width $>2,000$ km s$^{-1}$. The 322 type 1 AGNs in the \xcos\ sample were selected from a parent sample of 361 type 1 AGNs by excluding the 39 radio-loud AGNs identified using the same radio-loudness definition as described in \S\ref{subsec:bootes}.

As discussed in \S\ref{subsec:lxl6um_lir}, we utilize broad-band multiwavelength
photometry to determine the AGN contribution to the mid-IR luminosity
of our  AGNs. To this end, we also make use of the publicly available
broad-band photometry in the COSMOS survey region culled from
\cite{cat_cosmos07}, \cite{scosmoscatalog} and \cite{elvi12xmm}.
In detail, we first obtain the optical positions by cross-correlating the
XMM-COSMOS identification numbers (XIDs) of the \cite{luss10}
sources with those in the \xcos\ multiwavelength catalog \citep{brus10cosmosagn},
in which the optical to X-ray counterpart association is obtained based on a
likelihood ratio technique \citep[see \S3 of][for details]{brus10cosmosagn}.
We then cross-correlate the optical positions of the \cite{luss10} sources to
the \cite{cat_cosmos07} and \cite{scosmoscatalog} catalogs.
We use the broad-band photometry spanning optical to far-IR wavelengths that are
comparable to the \bootes\ survey region for the SED fits. In detail, we use the
{\it Subaru} optical photometry at {\it u,g,r,i,z} bands, and the near-IR {\it J, H, K} from
Calar Alto, UH 88'' and CFHT observatories, respectively. The mid-IR photometry
comes from both the {\it Spitzer} IRAC and MIPS instruments, including
3.6, 4.5, 5.8, 8.0, 24, 70, 160 $\micron$. Similar to Bo\"{o}tes, the COSMOS survey region is also covered by HerMES. Therefore, we also match the {\it Herschel} SPIRE photometry to the 322 {\it XMM}-COSMOS AGNs with a $5\arcsec$ search radius.
The detection fraction at SPIRE 250\micron\ for the 322 XMM-COSMOS type 1 AGNs is also $\sim 15\%$.
The $L_{\rm MIR}$ measurements and \lxlmir\ relation of these far-IR detected AGNs
are also discussed in more detail in \S\ref{subsec:lxl6um_lir} and in \S\ref{subsec:lxmir_host}.

\subsection{XRT-SDSS: Optical AGN from SDSS and {\it Swift}/XRT}\label{subsec:sdss}

\begin{figure*}
\hspace*{-0.4in}
\includegraphics[width=0.53\textwidth]{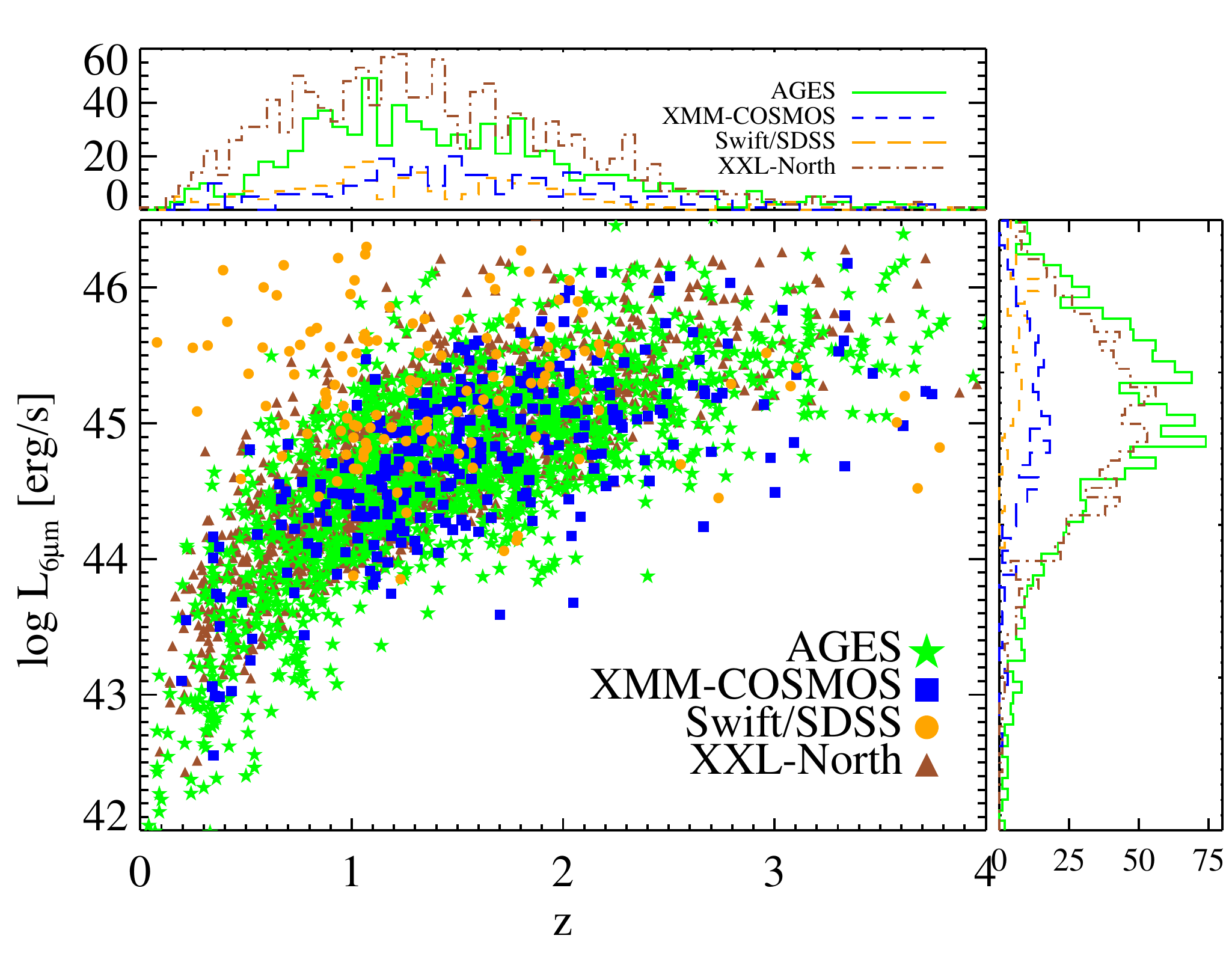}
\hspace*{0.1in}
\includegraphics[width=0.53\textwidth]{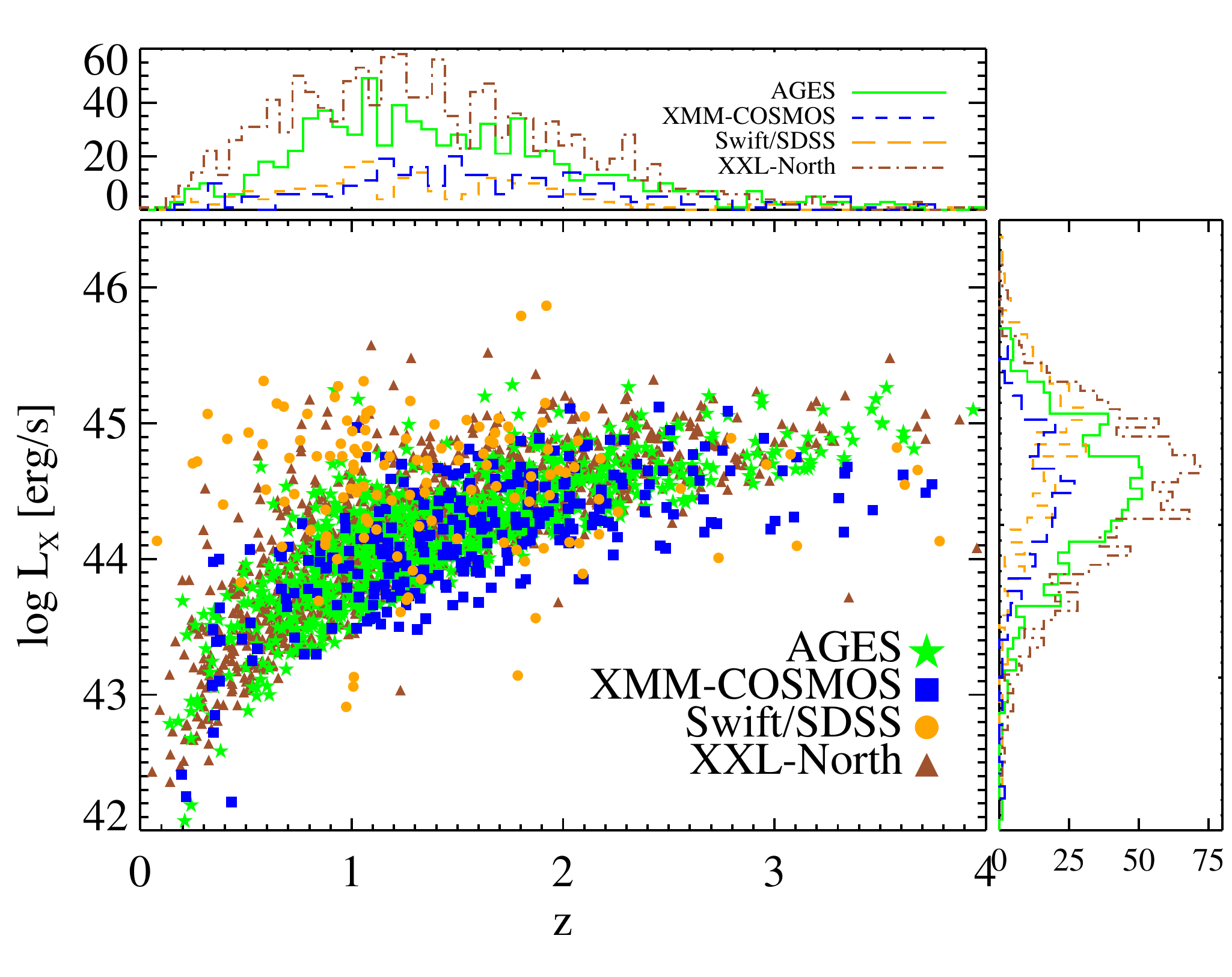}
\caption{The redshift distributions of $L_{6\micron}$ and $L_\mathrm{X}$ for
different samples studied in this work. This figure shows that our samples span
a wide range of X-ray and mid-IR luminosities, which is crucial for studying the
intrinsic \lxlmir\ relation of AGNs. The histograms of mid-IR and X-ray
luminosities for the different samples are shown in the top panels of the {\it left} and {\it right} figures, respectively.}
\label{fig:lz}
\end{figure*}

To investigate the \lxlmir\ relation, it is also
important to consider the possible biases created by missing the most luminous
sources due to the limited volume of surveys like \bootes\ and {\it XMM}-COSMOS.
For this purpose we use the {\it Swift}/SDSS catalog from \citet[][W12 hereafter]{wu12dr5}. \par

W12 matched all of the 77,429 optically selected SDSS DR5 quasars
\citep{sdssdr5,sdssdr5_quasar} to the {\it Swift}/XRT archive \citep{swift_xrt}
and found that there are 1,034 SDSS DR5 quasars within $20\arcmin$ of a {\it Swift}
pointing.
We refer to this catalog as the XRT-SDSS catalog throughout the rest of the paper.
Due to the serendipitous nature of the XRT-SDSS catalog, the {\it Swift}/XRT
exposure time ranges from $1-600$ ks. We follow the approach of W12 by focusing
only on the 607 objects that are unambiguously identified as quasars and excluded
objects that are radio-loud or obscured (see Table 9 of W12). W12 define a
``clean sample'' of quasars by enforcing a minimum XRT exposure time of $>10$ ks,
which includes 241 objects with a $82\%$ X-ray detection rate. To maximize the
sample size, we also consider a more liberal exposure time cut at $>5$ ks. The 5 ks
XRT-SDSS sample includes 362 objects with a 70\%  X-ray detection rate.

We also make use of the photometry from the SDSS DR5 quasar catalog by
cross-matching the SDSSID from the Table 7 of W12 to the SDSSID of the SDSS DR5
quasars catalog. All the quasars in the 10 ks and 5 ks samples have photometry
in the {\it u, g, r, i, z} bands and $\approx 18\%$ of the W12 quasars have 2MASS
{\it J, H} and {\it Ks} photometry. \par

To estimate the AGN mid-IR luminosity, we match the SDSS DR5
coordinates to the ALLWISE catalog using a matching
radius of $2\arcsec$.
We check the number of possibly misidentified sources by
randomly shifting the positions of the XRT-SDSS sources by 1 arcmin and matching
the shifted positions to the ALLWISE catalog. We find that $1.5\%$ of the
randomly shifted positions have a {\it WISE} counterpart within $2\arcsec$,
suggesting that the spurious matching rate between the XRT-SDSS sources and the
{\it WISE} catalog is about $1.5\%$, which has a negligible effect on the \lxlmir\ relation.
For the XRT-SDSS type 1 AGNs in the 10 ks and 5 ks samples, all of them have
detections in at least three {\it WISE} bands.

\subsection{XMM-XXL North X-ray AGN sample}\label{subsec:xxl}
The {\it XMM}-XXL-North survey (XXL-N hereafter) is the northern part of the {\it XMM}-XXL survey, which is comprised
of two separate $\sim 25$ deg$^{2}$ fields \citep{pier16xxl}. As part of the
SDSS-III survey, X-ray sources matched to SDSS photometric objects with {\it r}$ < 22.5$
in XXL-N were all targeted by SDSS-III's Baryon Oscillation Spectroscopic Survey
\citep[BOSS,][]{smee13_boss}. The spectroscopic and photometric properties of the
X-ray AGNs in XXL-N have recently been reported in \citet[][M16 hereafter]{menz16xxl}.
Of the 3,042 sources in the M16 catalog with BOSS spectra, 1,787 are classified
as ``broad-line'' AGNs based on the presence of broad emission lines
(H$\beta$, \ion{Mg}{2}, \ion{C}{3}, or \ion{C}{4}) with FWHM larger than 1,000 km s$^{-1}$.

For this work, we directly use the SDSS photometry and {\it WISE} photometry provided
by M16. The optical and mid-IR photometry in M16 is obtained by crossmatching
the {\it XMM} positions with the SDSS or {\it WISE} positions using a
likelihood-ratio matching method (see \citealt{geor11} and M16 for details).
To maximize the photometric coverage of this dataset, we also obtain 2MASS photometry
from the ALLWISE catalog, which provides the associations between the {\it WISE}
source and the closest 2MASS source within a 3\arcsec\ radius.

Notably, the M16 catalog includes X-ray sources from the XXL survey \citep{pier16xxl}
as well as the sources from the predecessor of XXL, the {\it XMM}-LSS survey.
The {\it XMM}-LSS survey is a $\approx 4.5$ deg$^2$  field at the center of XXL-N
with deeper {\it XMM}-Newton coverage (10--40 ks).
For this work, we consider only the sources with 0.5--2 keV fluxes above the
``completeness limit'' of the XXL survey, $5.0\times 10^{-15}$ ergs~s$^{-1}$~cm$^{-2}$.
This flux limit is equivalent to $\approx 1.1\times 10^{-14}$ \fluxcgs\ in the
0.5--7 keV band assuming a $\Gamma=1.8$ X-ray power-law spectrum.
With this flux limit, the number of broad-line AGNs is reduced to 1,372.
This approach also ensures that the XXL-N type 1 AGNs studied in this work have
uniform X-ray coverage and high mid-IR detection fractions in the {\it WISE} bands
($\sim 87\%$ for 3.4, 4.6 and 12$\micron$ bands).

We also match the XXL-N type 1 AGNs to the VLA FIRST catalog \citep{cat_first}
with a 2\arcsec radius. We calculate the radio-loudness of the 41 sources with
FIRST counterparts using the same approaches described in previous subsections.
Of the 41 sources with FIRST counterparts, 38 of them satisfy the same ``radio-loud''
definition and are excluded. Since our goal is to study the relation between X-ray
and mid-IR luminosities of type 1 quasars, we focus only on the 1,153 X-ray
detected type 1 AGNs that are not radio-loud and have a $>5\sigma$ detection
significance in at least three {\it WISE} bands.

\subsection{Key properties of samples}\label{subsec:sampleprop}
We list the key properties of the samples used in this work in
Table 1. A common feature of the sources selected from these four
catalogs is that the sources are all optically confirmed as broad-line
AGNs, which ensures our \lxlmir\ measurements should have a minimal impact due to obscuration.
The median {\it r}-band magnitudes of these samples are 21.0, 21.5, 19.0, 20.7 mag
for AGES, {\it XMM}-COSMOS, XRT-SDSS, and XXL-N, respectively.

Both the AGES and \xcos\ catalogs have heterogeneous spectroscopic depths.
AGES specifically targeted sources identified as an AGN at other wavelengths
down to {\it i} $<22.5$, while for other galaxies the limiting magnitude is
{\it i} $<20$ \citep[see][for details]{AGEScatalog}.
For the \xcos\ sample, the spectroscopic data come from existing SDSS spectra,
the magnitude-limited zCOSMOS catalog \citep[{\it i} $<22.5$,][]{lill09zcosmos},
and spectroscopic observations with MMT and IMACS/Magellan down to {\it i} $\approx 25$.
For the SDSS DR5 quasar catalog, the spectroscopic depth is brighter ($i\lesssim 19.1$
for low-redshift quasars and $i<20.2$ for higher redshift quasars, see \citealt[][for details]{sdssdr5_quasar}.)
As for the XXL-N X-ray AGN catalog, the spectroscopic depth of BOSS is {\it r}$<22.5$,
which is deeper than the SDSS DR5 quasar catalog and similar to \bootes\ and {\it XMM}-COSMOS.

For the \bootes\ and \swift\ samples, the mid-IR observations are from either {\it Spitzer}
or {\it WISE}, and are complete for these optically luminous AGNs\footnote{Only for the three shorter wavelength {\it WISE} bands.},
thus these samples are only flux-limited in the optical and X-ray bands.
Notably, the X-ray non-detected objects in the AGES and \swift\ samples are still
covered by X-ray observations, which allows us to take the X-ray non-detected sources
into account when measuring the $L_{\rm X}-L_{\rm MIR}$ relation.
Both the \xcos\ and XXL-N samples are X-ray selected, so these catalogs are flux-limited in the X-ray and optical wavelengths.
As for mid-IR observations, the {\it Spitzer} observations are complete for the \xcos\ sample and the {\it WISE} coverage for the XXL-N sample is also highly complete ($\sim 87\%$).

\section{Luminosities in the X-ray and mid-IR}\label{sec:lums}
In this section, we briefly describe the methods used to calculate the
rest-frame $2-10$ keV luminosity and the mid-IR luminosity for the AGNs in each catalog.
For comparison, we plot the redshift, X-ray luminosity
at rest-frame $2-10$ keV ($L_{\rm X}$), and the luminosity of the AGN
component at $6\;\micron$ ($L_{6\micron}$) in Figure~\ref{fig:lz}. The $L_\mathrm{X}$ and $L_{6\micron}$
histograms are also shown in Figure~\ref{fig:lz}.

\subsection{X-ray luminosity}\label{subsec:lx}
For the \bootes\ sample, the X-ray photon count rates in the 0.5-7 keV band
are converted to a flux using a conversion factor of $7.8\times 10^{-15}$ ergs cm$^{-2}$
s$^{-1}$ for an object with 4 counts in a 5 ks exposure.
This is derived based on the assumption of an
unabsorbed X-ray spectrum with a photon index of
$\Gamma=1.8$ \citep[see][for a complete discussion]{kent05,murr05}.
We then converted the 0.5-7 keV luminosity to the rest-frame 2-10 keV
luminosity with a {\it k}-correction using the same spectral index. We note that Galactic absorption column density for the \xbootes\ survey is negligible ($\approx 10^{20}$ cm$^{-2}$), but we still take it into account in our calculation for consistency with other samples.

For the \xcos\ sample, the rest-frame $2-10$ keV X-ray luminosity comes
from \cite{luss10}, in which the X-ray count rates in the $0.5-2$ keV
and $2-10$ keV bands are converted into rest-frame $2-10$ keV
luminosities using a Galactic column density $N_{\rm H} = 2.5 \times
10^{20}$ cm$^{-2}$ \citep[see][]{capp09xcos}, and assuming photon indicies of $\Gamma=2$ and
$\Gamma=1.7$ for the soft and hard bands, respectively.
For consistency, we revise the 2 - 10 keV X-ray luminosities using
the same approach and assumptions as applied to the \xbootes\ sample.
This causes changes of $\sim 8\%$ relative to the values in the \cite{luss10} due to the different assumptions regarding
the intrinsic photon-index of the X-ray power-law spectrum.
\par

For the \swift\ sample, the estimate of X-ray luminosity varies depending
on the quality of available X-ray data. We briefly describe the
approach taken by W12 and refer the readers to \S2 in \cite{wu12dr5}
for details. In the XRT-SDSS sample, the observed-frame $0.3-10$ keV flux
for each source with XRT counts $>100$ is derived by fitting
the observed counts to estimate the X-ray power-law index and intrinsic
absorption. For sources with between 30 and 100 XRT counts, the intrinsic absorption is fixed to zero while the spectral
index is still a free parameter. For sources with XRT counts less
than 30, the flux is obtained by assuming a fixed spectral index of
$\Gamma=2$ and a fixed zero intrinsic absorption.
For consistency, we again modify the derived $2-10$ keV $L_{\rm X}$ for the
XRT-SDSS sample to account for the different choices in the AGN intrinsic spectral index.
This causes changes of $\sim 15\%$ relative to the values reported in W12.

For the XXL-N sample, the X-ray luminosities come from \cite{liu16xxl} who estimated the {\it intrinsic} X-ray
luminosity by jointly fitting the {\it XMM}-Newton PN and MOS data
with the Bayesian X-ray Analysis package \citep[BXA,][]{soft_bxa}\footnote{We note
that one of the 1,153 AGNs is only detected at the 0.5--2 keV band. For this
object we calculate its 2--10 keV $L_{\rm X}$ based on the 0.5--2 keV flux
assuming a power-law X-ray spectrum with a $\Gamma=1.8$ photon index}.
The model used to fit the data is a combination of three different models that take the intrinsic power-law continuum, absorption, Compton scattering features, and a soft scattering component into account \citep[see \S4.1 of][for details]{liu16xxl}. To broadly assess if the BXA-based $L_{\rm X}$ are comparable to our other estimates of $L_{\rm X}$, 
we re-calculate the X-ray luminosities for the XXL-N sample by converting the 0.5--8 keV photon count rates reported in \cite{liu16xxl} to rest-frame 2--10 keV $L_{\rm X}$ assuming a power-law X-ray spectrum with a $\Gamma=1.8$ photon index. The average difference from the BXA estimates is only 0.03 dex.
This is not surprising, as type 1 AGNs X-ray AGNs have been found to have little to no absorption \citep[e.g.][]{hick07obsagn}. For this work, we adopt the BXA-based intrinsic $L_X$ to minimize the possible bias on the measured \lxlmir\ relation due to any X-ray absorption.

For the majority of type 1 quasars in AGES, {\it XMM}-COSMOS, and \swift\ samples, the X-ray absorption correction to the X-ray luminosity is not available from spectral fitting due to the limited photon counts. 
Therefore, the uncertainties in their X-ray luminosities were estimated based on Poisson noises of the count rates calculated using the \cite{gehr86} method.
For these luminous type 1 AGNs, the intrinsic X-ray absorption could be considered negligible. In particular, an X-ray stacking analysis has shown that the hardness ratios for type 1 AGNs in \bootes\ are consistent with little to no absorption \citep[$N_{\rm H} \sim 10^{20}$ cm$^{-2}$,][]{hick07obsagn}.
We discuss the possible effects of X-ray absorption on the
observed \lxlsix\ relation further in \S\ref{subsec:xabs}.

\subsection{SED fitting analysis and mid-IR luminosity}\label{subsec:lxl6um_lir}

To estimate the contamination from the host galaxy to the AGN mid-IR luminosity, we use SED-fits
to calculating the intrinsic, de-absorbed AGN mid-IR
luminosity, including {\it Herschel} far-IR photometry when available.

We follow the approach described in \cite{chen15qsosf} by fitting the
photometry with three different components: an AGN spanning
from near-UV to far-IR, a stellar population in the host galaxy and
a model for dust emission from reprocessed starlight. \par

We created {\it ad hoc} AGN templates by combining the near-UV to near-IR empirical AGN template from \cite{asse10sed} with the
infrared AGN SEDs from \cite{mull11agnsed} and
\cite{netz07qsosf}. For each AGN template, we create a grid of AGN templates with $0 < E(B - V) < 10$ using
a hybrid extinction curve combining an SMC-like (Small Magellanic Cloud) extinction
curve at $\lambda < 3300$ \AA\ \citep{gord98} with a Galactic extinction curve at longer wavelengths \citep{card89},
with $R_V = 3.1$ for both \citep[see][for details]{asse10sed}.

For the host galaxy templates, we consider two different
components: the contribution from the stellar population of the galaxy, which
accounts for the optical to near-IR emission; and a starburst
component, which represents the mid- to far-IR dust emission from
re-processed stellar light. For the stellar population component, we adopt the three empirical galaxy templates from \cite{asse10sed} representing starburst (Im),
continuous star-forming (Sbc) and old stars (elliptical), respectively.
We follow the approach described in \cite{chen15qsosf} by replacing the $>4.9\;\micron$ hot dust components of the Sbc and Im templates assuming a
Rayleigh-Jeans tail identical to the elliptical galaxy to create empirical stellar population templates without dust
emission. For the starburst component, we use a total of 171 starburst templates from \cite{ce01}, \cite{dh02}, and \cite{kirk12} to accommodate a wide range of spectral shapes of star-forming galaxies.
\par

Given the SED templates, we fit the observed
photometry using a $\chi^2$ minimization algorithm to find these best-fit SED for each object.
From the best-fit SEDs, we calculate the monochromatic luminosities of
the AGN component at $6\;\micron$.
To account for uncertainties in the derived $L_{6\micron}$ due to both
the uncertainties in the flux measurements and any degeneracy
between the AGN and host galaxy components, we employ a bootstrapping
approach. For each source, we randomly scatter the original
photometry in every band with their $1\sigma$ uncertainties and redo the SED-fits.
We repeat
this process 500 times for each source.
For the Bo\"{o}tes, \xcos\ , \swift\ , and XXL-N samples, the median $L_{6\micron}$
uncertainties are $0.08$, $0.11$, $0.12$ and $0.13$ dex, respectively.\par

For type 1 AGNs in our sample, the majority of the sources have mid-IR
SEDs dominated by the AGN component. The average AGN fraction (the
absorbed AGN component contribution at $6\;\micron$) for the \bootes\,
\xcos\ and \swift\  samples are $89\%$, $95\%$, $92\%$, and $81\%$,
respectively. For $98\%$ of the AGNs in our final sample, the AGN
component dominates (AGN fraction $>50\%$) at $6\;\micron$.
However, there is a caveat when estimating the host galaxy contamination at mid-IR wavelengths
with the SED fitting approach for the \swift\ sample. Unlike the
\bootes\ and the \xcos\ samples that include far-IR observations, the \swift\
sample relies on a very small number of {\it WISE} photometric data points to
constrain the host galaxy contribution in the mid-IR. Since starburst activity can also produce strong mid-IR
emission, we use the 228 far-IR detected sources in \bootes\ and
\xcos\ to examine whether the exclusion of far-IR photometry could
affect the observed AGN $6\;\micron$ luminosity.
We redid the SED-fits excluding the photometry at observed wavelengths longer than
$24\;\micron$ and then compared the estimates for $L_{6\micron}$. We found a median difference of only $\pm0.03$ dex, which is not surprising since the AGNs studied in this work are
luminous optical quasars. This 0.03 dex uncertainty was added in
quadrature to the measured $L_{6\micron}$ for all objects lacking far-IR
photometry.

\section{The Correlation between X-ray and mid-IR luminosities for X-ray detected quasars}\label{sec:xdet}
\begin{figure*}
\hspace*{-1.in}
\epsscale{1.1}
\plotone{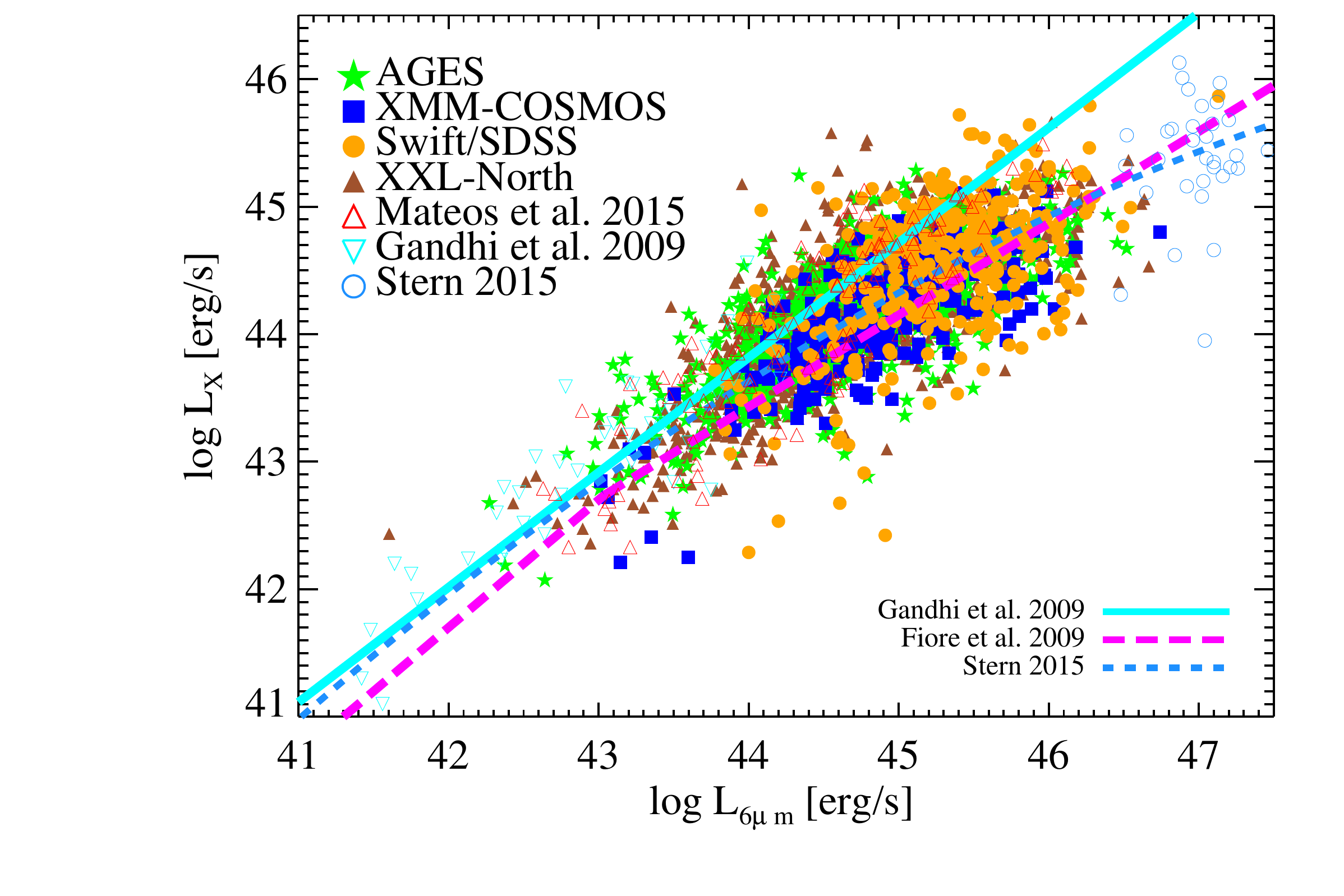}
\caption{The $L_\mathrm{X}-L_{6\micron}$ distribution for the four samples
studied in this work. For comparison, the type 1 AGN sample from \cite{mate15},
the Seyfert galaxies from \cite{gand09seyfir}, and the high luminosity SDSS quasars
from \cite{ster15} and \cite{just07} are also shown. The approximately linear
relation from \cite{gand09seyfir}, the luminosity dependent relation from \cite{fior09obsc},
and the luminosity dependent relation from \cite{ster15} are shown as the solid line,
long dashed line, and short dashed line, respectively. The
luminous AGNs in our samples have systematically lower X-ray luminosities than
predicted by the extension of the linear relation found for local Seyfert galaxies.}
\label{fig:lxl6umdata}
\end{figure*}

In Figure~\ref{fig:lxl6umdata}, we show the \lxlsix\ distributions of sources
with X-ray detections for all four catalogs.
For comparison, we also show the \cite{gand09seyfir}, \cite{mate15}\footnote{We note that for the five most X-ray luminous quasars in this sample, one of them is a lensed quasar and three of them are radio-loud. Therefore, we do not include these objects in this plot.} and \cite{ster15} samples, along with the $L_{\rm X}-L_{6\micron}$ relation for local Seyfert galaxies
from \cite{gand09seyfir}. Clearly, the four catalogs studied in this work show a $L_{\rm X}-L_{6\micron}$ distribution departing from the \cite{gand09seyfir} relation and other roughly linear
relations suggested by studies of local active galaxies
\citep[e.g.,][]{lutz04irx,maio07,asmu15}. The
$L_{\rm X}-L_{6\micron}$ distributions for our samples are in
broad agreement with the \cite{fior09obsc} and \cite{ster15} luminosity-dependent \lxlsix\ relations for luminous X-ray AGNs. We have converted the monochromatic luminosities measured at different wavelengths for these comparison samples (e.g. $5.8\micron$ and $12\micron$) to $L_{6\micron}$ using the \cite{asse10sed} AGN template.
\par

\begin{figure*}
\hspace*{-1.in}
\epsscale{1.1}
\plotone{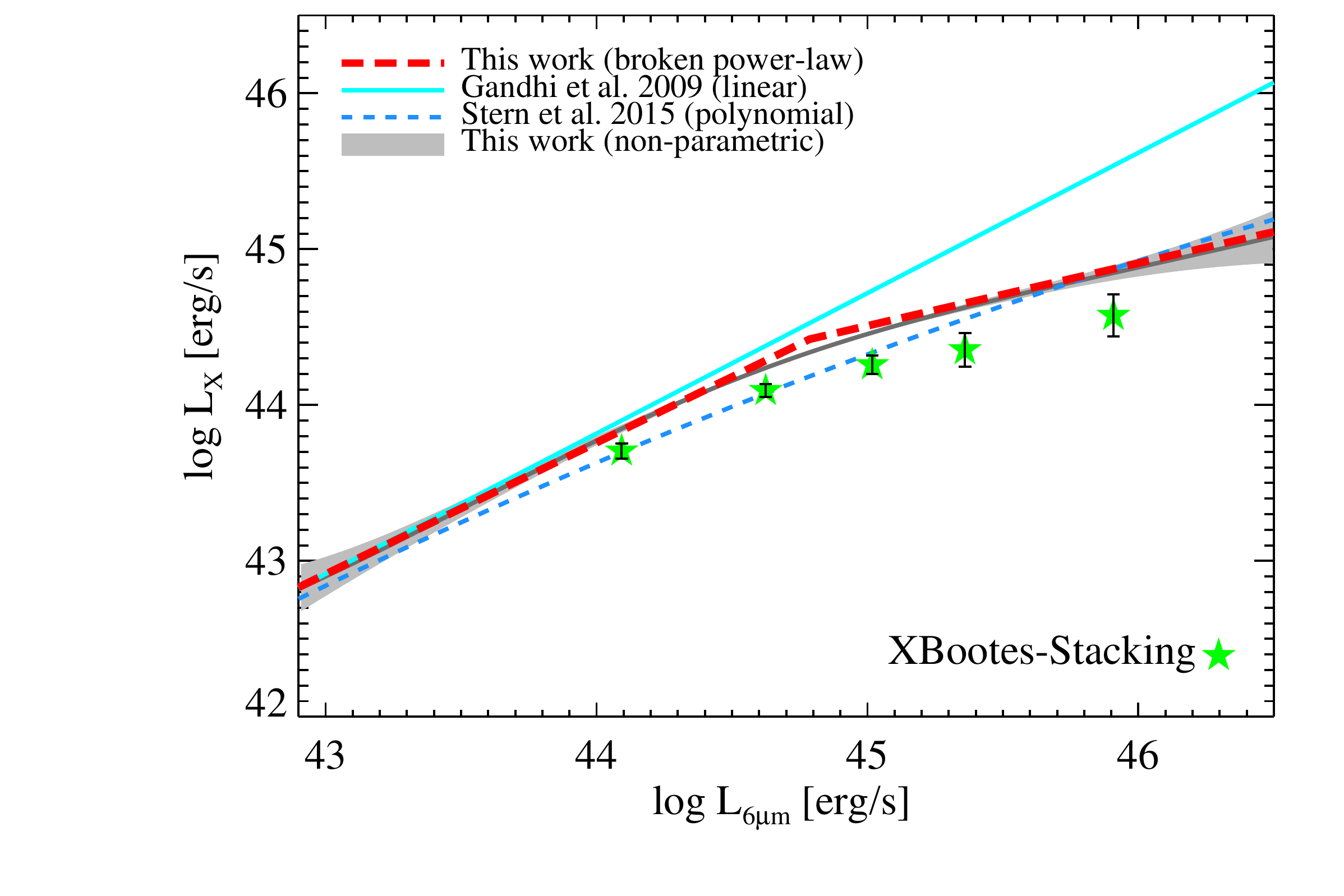}
\caption{Updated \lxlsix\ relations derived using the X-ray detected type 1 AGNs studied in this work. The updated relation is best-described by a broken power-law (red, thick dashed line). At high $6\micron$ luminosities, our result is more consistent with the \cite{fior09obsc} and \cite{ster15} luminosity-dependent \lxlsix\ relations than the \cite{gand09seyfir} linear relation. The non-parametric fit to the X-ray detected sources is also shown as the dark-gray line with the $1\sigma$ uncertainty indicated by the shaded region. We also include the \xbootes\ X-ray stacking results to show the effect of X-ray non-detections as the green stars.}
\label{fig:lxl6um_relations}
\end{figure*}

To obtain a simple parametrized \lxlsix\  relation for AGNs spanning a wide
range of AGN luminosity, we fit the combined Bo\"{o}tes, {\it XMM}-COSMOS,
XRT-SDSS, and XXL-N samples assuming that their \lxlsix\ relation
follows the bilinear equation 

\begin{equation}\label{eq:bpl}
\begin{aligned}
\log L_{6\micron} & < \log L_{6\micron}^\star:\\
\log L_X & =m_1\times\log \frac{L_{6\micron}}{10^{45}\mathrm{erg\;s}^{-1}}+b_1\\
\log L_{6\micron} & \geq \log L_{6\micron}^\star:\\
\log L_X & = m_2\times\log \frac{L_{6\micron}}{10^{45}\mathrm{erg\;s}^{-1}}+b_2,
\end{aligned}
\end{equation}
discussed by \cite{fior09obsc}.
Here $L_{6\micron}^\star$ is the ``breaking luminosity'',
$(m_1, b_1)$ and $(m_2, b_2)$ stand for the slope and intercept for each segment of the bilinear relation.
This equation is identical to the assumption that $L_\mathrm{X}$ and $L_{6\micron}$ follow a
broken power-law relation in the linear space. Since the broken power law relation
assumes continuity on the breaking point, the number of free parameters is 3,
because $b_2=\log L_{6\micron}^\star\times(m_1-m_2)+b_1$.
We next fit the data using an iterative $\chi^2$ minimization algorithm (Levenberg–Marquardt) based on the {\sc MPFIT} package in {\sc IDL}.
The best-fitting parameters for ($\log L_{6\micron}^\star$, $m_1$, $m_2$, $b_1$, $b_2$) are
$(44.79,0.84,0.40,44.60,44.51)$, and the corresponding uncertainties are
$(0.11,0.03,0.03,0.01,0.01)$. We show this best-fit broken power-law relation in
Figure~\ref{fig:lxl6um_relations}.
The break luminosity of our bilinear relation is significantly higher
than that found by \cite{fior09obsc}. This is not surprising, as
there are very few objects in the \cite{fior09obsc} sample with mid-IR luminosity
smaller than their break luminosity.
For our combined sample of X-ray detected type 1 AGNs, there are 1,301 sources
with $L_{6\micron}$ smaller than our best-fit break luminosity,
$L_{6\micron} = 6.2\times 10^{44}$ ergs s$^{-1}$. 
We also fit the data with a simple linear relation, $\log L_X =\alpha\times\log \frac{L_{6\micron}}{10^{45}\mathrm{erg s^{-1}}}+\beta$, and an f-test rejects this model over the bilinear model with a $>99.9\%$ confidence level according to the f-test probability. 
It is also important to note that the location of the break luminosity might depend
strongly on how the sample populates the \lxlsix\ parameter space. 
To assess how the sparse distribution of our sample in \lxlsix\ affects the results, 
we divide our sample into six $L_{6\micron}$ bins of approximately equal size. 
We then weight the total $\chi^2$ of each bin by its source number such that
$L_{6\micron}$ bins with smaller source numbers have similar statistical 
power to the $L_{6\micron}$ bins with larger source numbers. 
We find that the break luminosity increases by 0.5 dex with this approach.

In practice, the result of fitting a bilinear \lxlsix\ relation to an unevenly
distributed AGN sample will not only depend on the intrinsic \lxlsix\ slopes, but also
the relative numbers of low-luminosity AGNs and luminous quasars.
Due to the volume and flux-limited nature of extragalactic surveys, it is extremely difficult to construct a sample that could populate the \lxlsix\ parameter space as evenly as our simple weighted $\chi^2$ minimization exercise effectively does. Therefore, we consider the result of this simple exercise an ``upper limit'' on the break luminosity of a bilinear \lxlsix\ relation, and conclude that the linear \lxlsix\ relation for lower luminosity AGNs cannot be extended to quasars that are more luminous than $L_{6\micron} = 1.4\times 10^{45}$ ergs s$^{-1}$.

In Figure~\ref{fig:lxl6um_relations}, we compare our bilinear regression fit with the \cite{gand09seyfir} linear relation for local Seyferts and the second order polynomial fit of \cite{ster15}.
While our bilinear relation is largely consistent with the second order polynomial fit of \cite{ster15}, it is not clear which regression model is the best option. Therefore, in addition to assuming that the \lxlsix\ distribution follows a specific functional form, it is also useful to adopt a non-parametric approach to
visualize the relation between $L_\mathrm{X}$ and $L_{6\micron}$ of our sample.

We use the Gaussian process regression algorithm from {\sc Gpy}\footnote{\url{http://sheffieldml.github.io/GPy/}}with a polynomial kernel function to fit all of the X-ray detected sources studied in this work. The Gaussian process regression
algorithm assumes that the dependent variable (the $L_{\rm X}$ in this work) of a sample of finite size could be described by a multivariate Gaussian function with dimensions equal to the sample size. With an assumed kernel function (i.e. the covariance function of the Gaussian process), the Gaussian process algorithm also takes the measurement uncertainties into account and analytically finds the best-fit multivariate Gaussian function. We can then compute the non-parametric prediction using the posterior probability function based on the Gaussian process regression results.

In Figure~\ref{fig:lxl6um_relations}, we also show this non-parametric fit to
the data with 1$\sigma$ uncertainty as the
gray shaded region. We find that both our bilinear regression fit and the second
order polynomial fit of \cite{ster15} follow the non-parametric prediction closely,
which demonstrates that a change of the \lxlsix\ slope occurs at
$L_{6\micron}\sim 10^{44}$--$10^{45}$ $\textmd{erg}\;\textmd{s}^{-1}$.

\section{Possible biases and comparison to the literature}\label{sec:bias}

In this section, we explore how the different observational constraints affect the estimate of the intrinsic \lxlsix\ relation for AGNs. For simplicity, we do not consider the broken power-law regression when addressing various observational constraints. Instead, we focus on how X-ray absorption, X-ray non-detections, and X-ray flux limits affect the linear \lxlsix\ relation of the more luminous objects in our sample (i.e., the quasars with $\log L_{6\micron}/$erg s$^{-1} > 43.8$, which corresponds to a bolometric luminosity of $10^{45}$ erg s$^{-1}$ assuming the \citealt{hopk07qlf} bolometric correction factors).

\begin{table*}
\scriptsize
\begin{center}
\setlength{\tabcolsep}{1.5mm}
\caption{Best-fit parameters for $L_{\rm X}=\alpha\times L_{6\micron}+\beta$ calculated using the \cite{kell07} method}\label{tbl:lxl6umpars}
\begin{tabular}{lchcccc}
\tableline\tableline
\multicolumn{7}{c}{} \\[-0.5ex]
\multicolumn{1}{c}{Sample} &
\multicolumn{1}{c}{Description} &
\multicolumn{1}{c}{} &
\multicolumn{1}{c}{$N_\mathrm{XD}$} &
\multicolumn{1}{c}{$N_\mathrm{UL}$} &
\multicolumn{1}{c}{$\alpha$} &
\multicolumn{1}{c}{$\beta$} \\[1ex]
  \multicolumn{1}{c}{} &
  \multicolumn{1}{c}{} &
  \multicolumn{1}{c}{} &
  \multicolumn{1}{c}{(1)} &
  \multicolumn{1}{c}{(2)} &
  \multicolumn{1}{c}{(3)} &
  \multicolumn{1}{c}{(4)} \\[1ex]
\tableline
\multicolumn{5}{c}{I. Best-fit parameters for luminous quasars (see \S\ref{subsec:lxl6um_xnd})} \\
\tableline
Bo\"{o}tes$^a$          & Stacking                                &              & 727    & 620$^b$ & $0.51\pm0.06$ & $44.23\pm0.05$  \\
Bo\"{o}tes          & X-ray detected                          & $>10^{43.8}$ & $727$      & $0$     & $0.50\pm0.02$ & $44.36\pm0.01$ \\
Bo\"{o}tes          & All                                     & $>10^{43.8}$ & 727        & $620$   & $0.49\pm0.03$ & $44.02\pm0.02$ \\
{\it XMM}-COSMOS    & All, X-ray selected                     & $>10^{43.8}$ & 293        & 0       & $0.51\pm0.03$ & $44.30\pm0.02$ \\
{\it Swift}/SDSS (10 ks) & X-ray detected                     & $>10^{43.8}$ & 198        & 0       & $0.56\pm0.05$ & $44.31\pm0.03$\\
{\it Swift}/SDSS (10 ks) & All                                & $>10^{43.8}$ & 198        & 43      & $0.56^{+0.05}_{-0.06}$ & $44.24\pm0.03$\\
{\it Swift}/SDSS (5 ks)  & X-ray detected                     & $>10^{43.8}$ & 247        & 0       & $0.58^{+0.05}_{-0.04}$ & $44.40\pm0.03$\\
{\it Swift}/SDSS (5 ks)  & All                                & $>10^{43.8}$ & 247        & 115     & $0.58\pm0.05$ & $44.24\pm0.03$\\
{\it XMM} XXL-N              & X-ray selected                     & $>10^{43.8}$ & 1071       & 0       & $0.59\pm0.02$ & $44.36\pm0.01$\\
\\
\tableline\tableline
\multicolumn{5}{c}{II. Effects of X-ray absorption (see \S\ref{subsec:xabs})} \\
\tableline
\vspace{0.5ex}
Bo\"{o}tes (X-ray detected)& Assuming 20\% of the sample are X-ray obscured & $>0$ & 727& 0 & $0.49\pm0.05$ & $44.41\pm0.01$ \\
\\
\tableline\tableline
\multicolumn{5}{c}{III. Effects of X-ray flux limit (see \S\ref{subsec:lxl6um_fl})} \\
\tableline
\vspace{0.5ex}
Mateos et al. 2015             & Type 1 AGNs with $L_{6\micron}>10^{43.8}$ erg s$^{-1}$ &      & 103& 0 & $0.81\pm0.06$ & $44.58\pm0.04$\\
Bo\"{o}tes (X-ray detected)& $f_X>1.0\times10^{-13}$ [erg/s] (M15)                  & $>0$ & 31 & 0 & $0.80^{+0.11}_{-0.12}$& $44.59\pm0.07$ \\
Bo\"{o}tes (X-ray detected) & $f_X>5\times10^{-14}$ [erg/s] (W12 5ks)    & $>10^{43.8}$ & 163 & 0 & $0.58\pm0.05$ & $44.45\pm0.03$ \\
Bo\"{o}tes (X-ray detected) & $f_X>2.5\times10^{-14}$ [erg/s] (W12 10ks) & $>10^{43.8}$ & 403 & 0 & $0.52\pm0.03$ & $44.42\pm0.02$ \\
Bo\"{o}tes (X-ray detected) & $f_X>1.1\times10^{-14}$ [erg/s] (XXL-N) & $>10^{43.8}$ & 665 & 0 & $0.51\pm0.02$ & $44.38\pm0.01$ \\
\tableline\tableline
\end{tabular}
\end{center}
{\bf Note:--}
$^a$ Best-fit parameters for the \bootes\ stacking results are derived using a
simple $\chi^2$ minimization method, see \S\ref{subsec:lxl6um_xnd}.\\
$^b$ The number of stacked sources.
$(1)$ Number of X-ray detected quasars.
$(2)$ Number of quasars with only an $L_X $ upper limit.
$(3)$ Slope of the best-fit \lxlsix\ relation.
$(4)$ Intercept of the best-fit \lxlsix\ relation in $\log$ ergs s$^{-1}$.\\
Part I: The best-fit parameters for different samples. See \S\ref{subsec:lxl6um_xnd} for a complete discussion.\\
Part II: The median value of the best-fit parameters between the intrinsic $L_{\rm X}$ and $L_{6\micron}$ assuming $20\%$ of the \bootes\ sample are heavily X-ray obscured. We show that the parameters do not change significantly. See \S\ref{subsec:xabs} for details.\\
Part III: The best-fit parameters of the \bootes\ subsamples selected with different flux limits. See \S\ref{subsec:lxl6um_fl} for a complete discussion.
\end{table*}

\subsection{Accounting for X-ray non-detected type 1 quasars}\label{subsec:lxl6um_xnd}
For the shallow flux limits of the \bootes\ and XRT-SDSS samples, a significant fraction of sources are not detected in
the X-rays.  For the \bootes\ type 1 AGNs, \cite{hick07obsagn} used an X-ray stacking analysis to show that the X-ray
properties of the X-ray undetected type 1 mid-IR quasars are
consistent with luminous X-ray AGNs with little or no absorption.
Therefore, it is important to take into account the average X-ray contribution for those optically
unobscured quasars without a direct X-ray detection when deriving the
\lxlsix\  correlation since their average X-ray luminosity could be
non-negligible. \par

For the \bootes\ quasars, we used an X-ray stacking analysis to
account for sources not individually detected in X-rays. We divide
these quasars into five bins of $L_{6\micron}$ and calculated their stacked X-ray luminosity. We define the stacked X-ray counts as the average number of
background-subtracted photons detected within the 90\% point-spread
function (PSF) energy encircled radius at 1.5 keV, $r_{90}$, where
$r_{90}=1''+10''(\theta/10')^2$ and $\theta$ is the angle
from the {\it Chandra} optical axis\footnote{{\it Chandra} Proposers'
  Observatory Guide (POG), available at
  \url{http://cxc.harvard.edu/proposer/POG}.}. We only include sources within $\theta < 6\arcmin$ ($r_{90}<4\farcs6$) in the analysis.
We adopt background surface brightnesses of 3.0 and 5.0 counts s$^{-1}$ deg$^{-2}$ for the
$0.5-2$ keV and $2-7$ keV bands, based on the estimates of the diffuse
background from \cite{hick07obsagn}. We convert count rates (counts
s$^{-1}$) to flux (ergs~s$^{-1}$~cm$^{-2}$) using the conversion
factors $6.0\times10^{-12}$ ergs cm$^{-2}$ count$^{-1}$ in the 0.5-2
keV band and $1.9\times10^{-11}$ ergs cm$^{-2}$ count$^{-1}$ in the
2-7 keV band.  To estimate the average X-ray stacking luminosity from
the X-ray flux, we assume that all galaxies in each bin of
$L_{6\micron}$ reside at the average luminosity distance for the
galaxies in that bin. More details of the stacking procedure are
described in \S 5.1 of \cite{hick07obsagn}. \par

To derive the \lxlsix\ relation including both the X-ray detected and non-detected sources, we calculate the average $L_\mathrm{X}$ in bins of $L_{6\micron}$ by taking the weighted average of the individually
detected sources and the stacking luminosity. We find that the
average $L_X$ of the entire \bootes\ sample has an $\lx\
-\lsix\ $ slope similar to that of the X-ray detected sources derived
in \S\ref{sec:xdet} (also see Figure~\ref{fig:lxl6um_relations}) with a
smaller intercept:
$\log L_X = (0.51\pm0.06)\times\log L_{6\micron}/10^{45}\mathrm{erg\;s^{-1}}+(44.23\pm0.05)$.\\

Another useful approach to take the X-ray non-detected sources into account when deriving the linear relation between $L_\mathrm{X}$ and $L_{6\micron}$ is the Bayesian maximum likelihood method presented by \citet[][K07 hereafter]{kell07}. The K07 method determines the best-fit linear relation by sampling the $L_\mathrm{X}$ values of
non-detected sources from the prior provided by the detected sources
and the value of the upper limits. For the \bootes\ and the XRT-SDSS samples,
the $L_{\rm X}$ upper limits for the X-ray non-detected sources were calculated using the flux limits and the corresponding redshift for each source. We then use the
K07 method to recalculate the best-fitting parameters for the complete
\bootes\ and XRT-SDSS samples, respectively. In detail, we use K07's method to perform Markov Chain Monte Carlo simulation using a Metropolis–-Hastings algorithm sampler with 10,000 iterations to obtain the posterior probability distribution of the linear regression parameters. The best-fitting parameters are determined as the median of the posterior probability distributions of the model parameters. We adopt the 1$\sigma$ (68\%) uncertainties as the 16 and 84 percentiles of the posterior probability distributions. \par

In the first part of Table~\ref{tbl:lxl6umpars}, we list the best-fitting
parameters for different samples with and without the consideration of X-ray
non-detections calculated using the K07 method. For comparison, we also list the best-fit
minimum $\chi^2$ regression result for the data from \bootes\ stacking analysis.
We also list the best-fit parameters for \xcos\ and XXL-N samples calculated using the K07 method.
For the \swift\ sample, due to the luminous nature of the SDSS DR5 catalog and the 10ks X-ray exposure time cut, the X-ray detection fraction is  $>83\% $. This leads to a similar \lxlsix\  relation regardless of the treatment of the X-ray non-detections. For the \bootes\ sample, we find that the inclusion of X-ray
non-detected sources does lead to a best-fit with a smaller $L_\mathrm{X}$ throughout the $L_{6\micron}$
range. This might be due to Eddington bias, as the \xbootes\ sources have as few as 4 counts in the 5 ks \xbootes\ {\it Chandra} observations.
The slopes of the \lxlsix\ relation also do not change
significantly when compared to the best-fit result for the X-ray
detected sources. The slopes derived from the stacked average are also consistent with the results
using the K07 regression analysis on samples with and without the
inclusion of X-ray non-detections.

The results suggest that the inclusion of X-ray non-detected objects does not alter the \lxlsix\ relation significantly given the relatively high X-ray detection fraction of the \bootes\ and {\it Swift}/SDSS samples.

\subsection{Effects of X-ray absorption on \lxlsix\ }\label{subsec:xabs}
As we mentioned at the end of \S\ref{subsec:lx}, a large fraction of our type 1
quasars do not have sufficient X-ray photon counts to properly
correct for attenuation of the observed $L_\mathrm{X}$ due to the small
gas absorption. The recent study of \lxlsix\ \citep{ster15} using the type 1 quasars from \cite{just07} also assumed that type 1 quasars have negligible X-ray absorption column densities.
While the average hardness ratio for type 1 AGNs does support this assumption \citep{hick07obsagn}, a recent study by \cite{merl13} 
that also focused on the XMM-COSMOS type 1 AGNs has also shown that a non-negligible fraction ($\sim 20\%$) of optical type 1 AGNs have a hardness ratio consistent with $N_{\rm H}>10^{22}$cm$^{-2}$ \citep[see Figure 4 of][]{merl13}.

Although the effect of absorption is less significant for high redshift quasars 
because the observed X-rays correspond to harder rest-frame energies, 
it is still important to understand the impact of gas absorption on the observed \lxlsix\ relation.
We estimate the effect of the bias caused by the possible presence of X-ray obscuration by conservatively
assuming that $20\%$ of the sources have $N_{\rm H}=10^{23}$ cm$^{-2}$.  We note that this assumption is an extreme case in which many of the optical type 1 AGNs are heavily obscured in the X-ray. For the {\it XMM}-COSMOS sample studied by \cite{merl13}, less than $5\%$ of the optical type 1 AGNs have hardness ratios consistent with $N_{\rm H}>10^{23}$ cm$^{-2}$.
We estimate the correction on the \lxlsix\ relations using the following steps:  (1) Randomly select $20\%$
of the X-ray detected sources. (2) Assume these objects have an absorption column density of $N_{\rm H}=10^{23}$ cm$^{-2}$, and  calculate the ``de-absorbed'' X-ray luminosity with an intrinsic X-ray spectral index of $\Gamma=1.8$.
(3) Recalculate the best-fit \lxlsix\ relations using the de-absorbed X-ray luminosities and the K07 method.
(4) Repeat (1) through (3) for 1,000 times.
The result of this bootstrapping analysis suggests that even for the unlikely case in which $20\%$ of the optical type 1 AGNs have heavy absorbing column densities, the slope of \lxlsix\ is only $\approx 0.01$ lower than 
the result neglecting absorption. The median values and standard deviations of the slopes and intercepts from the bootstrapping analysis are listed in Table~\ref{tbl:lxl6umpars}.

A larger part of this is that Compton-thin absorption has little effect on X-ray luminosity estimates for higher redshift sources (e.g., see \citealt{just07} and \citealt{ster15}). At $z\sim 1$, even for an AGN obscured by $N_{\rm H}\sim 10^{23}$ cm$^{-2}$ column densities, its observed-frame 0.5-7 keV flux would only be attenuated by $\sim 50\%$. Therefore, it is not surprising that the \lxlmir\ relation does not change significantly even for the case in which $20\%$ of the type 1 AGNs are X-ray absorbed. 
Since gas absorption only has a small effect on the slope of the \lxlmir\ relation, the difference between the flattened \lxlsix\ relation found in our study (and \citealt{fior09obsc}, \citealt{ster15}) and the linear relation that extends to luminous AGNs reported by \cite{mate15} and \cite{asmu15} must be caused by factors other than the possible presence of absorption in optical type 1 AGNs. 

\subsection{Effects of X-ray flux limits}\label{subsec:lxl6um_fl}

\begin{figure*}
\hspace*{-0.5in}
\plotone{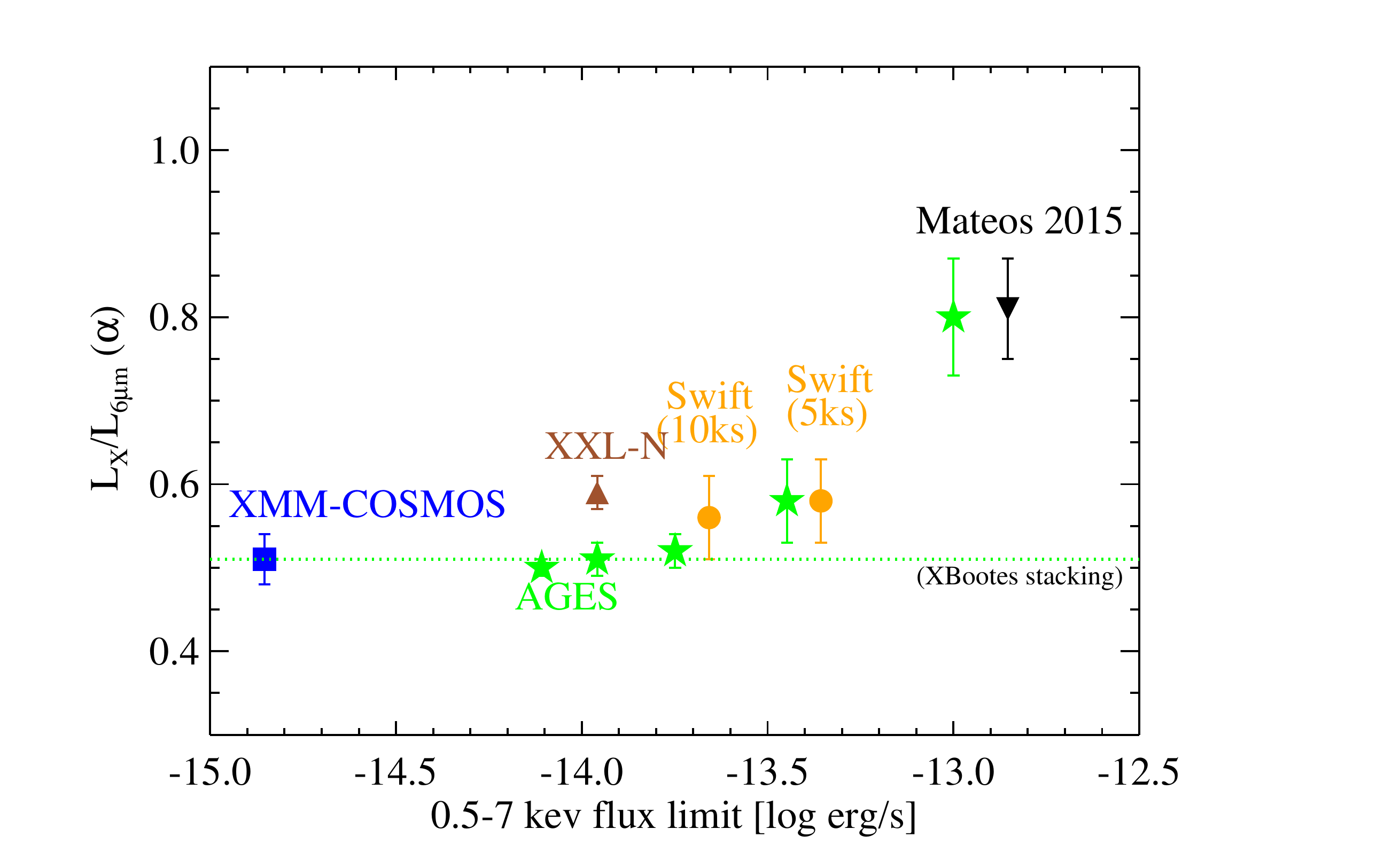}
\caption{The dependence of the slope of \lxlsix\  relation on the effective X-ray flux limit. The green stars are the  $L_X/L_{6\micron}$ slopes for X-ray detected subsamples from AGES selected with different X-ray flux limits. For comparison, the M15 and \swift\ samples from which the X-ray flux limits are drawn from are also plotted as the downward triangle (M15), and the yellow circles (\swift\ ).
The \xcos\ sample is shown as the blue squares.
We note that all of the $L_{\rm X}/L_{6\micron}$ slopes are derived for the luminous ($L_{6\micron}>10^{43.8}$ \ergs\ ) quasars using the K07 method.}
\label{fig:xfl}
\end{figure*}

As we have shown in \S\ref{subsec:lxl6um_xnd}, the exclusion of X-ray non-detected sources could result in a biased \lxlsix\ relation, but the effect is within the uncertainty for the \bootes\ and \swift\ samples due to their high X-ray detection fractions.
In \S\ref{subsec:xabs}, we also show that X-ray absorption should have little effects on the \lxlsix\ relation for luminous quasars.
Here we examine if the X-ray flux limits are the primary factor that drives the various \lxlsix\ relations reported in the literature.
We note that the mid-IR survey flux limits for luminous quasars are more homogeneous across different studies than the X-ray flux limits, as almost every luminous optical quasar in studies of the \lxlsix\ relation have clear detections at mid-IR wavelengths.

Recently, \citet[][M15 hereafter]{mate15} reported a
$L_X-L_{6\micron}$ relation for the hard X-ray ($4.5-10$ keV) sample
selected from Bright Ultra-hard \xmm\ Survey (BUXS). They found
an approximately linear $L_X-L_{6\micron}$ relation even for X-ray luminous quasars with $L_{\rm X}$ up to $\approx 10^{46}$ erg s$^{-1}$, in disagreement with our results and the results from \cite{fior09obsc,ster15}.\footnote{M15 fitted the SEDs of their sources with separate AGN accretion disk and AGN torus
components. Thus the $L_{6\micron}$ in M15 is inevitably lower than the $L_{6\micron}$ of our
work, which is derived by decomposing the AGN and host galaxy
components instead of separating the AGN accretion disk and AGN torus
components. However, we note that the difference is small as the typical AGN
accretion disk contribution is very small at $6\;\micron$ ($\sim 9\%$,
according to M15). In this work, we compare the $L_X-L_{6\micron}$
relation from M15 with our results by adding the average $9\%$ AGN accretion disk emission back to the $L_{6\micron}$ reported in that work.}
The M15 sample has a hard ($4.5-10$ keV) X-ray flux limit of $6\times10^{-14}$ erg
s$^{-1}$cm$^{-2}$. To convert the flux limit to an energy range
comparable to the $0.5-7$ keV of the {\it Chandra} observations, we assume a simple power-law X-ray spectrum with a photon index
$\Gamma=1.8$. Thus, the flux limit for the BUXS is equivalent to
$\sim 1.6\times10^{-13}$ erg s$^{-1}$cm$^{-2}$ at $0.5-7$ keV, which is
approximately 2 dex shallower than the {\it XMM}-COSMOS sample and 1
dex shallower than the \bootes\ and the XRT-SDSS samples.

To test whether the different slope observed in M15 is due to the
shallow X-ray flux limit, we apply several different X-ray flux limits to the \bootes\
sample and examine their effect on the slope of
the derived \lxlsix\ relation.
We first apply the \swift\ 10ks and 5ks flux limits
and the XXL-N flux limit
to the \bootes\
sample, and find that the $L_X/L_{6\micron}$ slopes for the \bootes\ sample
decrease for lower flux limits.

We next apply the converted BUXS flux limits to the \bootes\ sample and recalculate the \lxlsix\ relation using
the K07 method. Due to the shallowness of the BUXS flux limit, there
are only 38 sources in \bootes\ with an X-ray flux larger than
$1\times10^{-13}$ erg s$^{-1}$cm$^{-2}$.
Considering the difference in survey area, the number of sources is consistent with the M15
type 1 sample.
For the \bootes\ type 1 AGNs, only 4\% of the sources would have a flux limit
higher than that of the M15 sample, suggesting that surveys with shallow X-ray flux limits 
wll produce a biased \lxlsix\ relation because they miss the vast majority of the AGN population of similar $L_{\rm MIR}$.

The best-fit \lxlsix\  slope estimated using the K07 method for the
\bootes\ subsample with the $1\times10^{-13}$ erg s$^{-1}$cm$^{-2}$ cut is significantly steeper than the original \bootes\ sample.
This highlights the necessity of deep X-ray observations in order to reveal the intrinsic \lxlsix\ relation when the X-ray survey flux limits are too shallow.
Figure~\ref{fig:xfl} shows the effect of survey X-ray flux limits on the slope of the \lxlsix\ relation.

It is also interesting that the best-fit slope for the XXL-N sample is $0.59\pm0.02$, which is higher than that of the X-ray detected AGNs in Bo\"{o}tes, $0.50\pm0.02$ despite their similar X-ray flux limits. 
This is likely due to the fact that XXL-N has more extremely luminous sources due to its larger survey volume. 
It is also possible that a small number of high redshift radio-loud quasars in XXL-N were not identified in the shallow VLA FIRST catalog ($\sim 1$ mJy). 
Nonetheless, the best-fit \lxlsix\ slopes for the four samples studied in this work are still much smaller than the results reported in previous studies with much shallower flux limits, suggesting that the intrinsic \lxlmir\ relation could only be recovered by considering samples selected from both deep and wide X-ray surveys.

\subsection{Host galaxy contamination at mid-IR wavelengths}\label{subsec:lxmir_host}
In contrast to X-ray flux limits that might cause a steeper $L_X/L_{6\micron}$ relation,
contamination from host galaxies at mid-IR wavelengths could cause the observed $L_X/L_{6\micron}$ relation to be shallower than its intrinsic value \citep[e.g.,][]{luss13}.

We have carefully modeled the possible cool dust contamination 
using the strong constraints provided by {\it Herschel} observations at far-IR wavelengths. 
Here we further scrutinize the SEDs for the X-ray non-detected, mid-IR bright type 1 quasars in our \bootes\ and \xcos\ samples. We find that they have a median AGN fraction of $88\%$ at rest-frame $6\;\micron$, suggesting that their mid-IR SEDs are almost entirely dominated by the AGN component.
For the high $L_{6\micron}$ sources in our
sample, the W3 and MIPS $24\;\micron$ bands still show no
signs of strong PAH emission or silicon absorption.
While it is possible for strong nuclear starbursts to produce mid-IR continuum
with spectra similar to that of the AGN template
\citep[e.g.,][]{ball08nsb}, local nuclear starburst galaxies are often
hosted by less massive galaxies with moderate luminosity AGNs with
$L_{\rm X}$ less than the average $L_{\rm X}$ of our mid-IR bright
quasars. In \S\ref{subsec:lxl6um_lir}, we also demonstrated that the
host galaxy contamination is small even for far-IR luminous objects in
\bootes\ and {\it XMM}-Newton.
Thus, we argue that for our sample of luminous quasars,  it is
unlikely that the high $L_{6\micron}$ derived from the best-fitting
AGN template is due to substantial contamination from their host
galaxies. Since the mid-IR luminosity in our type 1 quasars is indeed
powered by the AGN, the shallower \lxlsix\ slope suggests that the AGN
X-ray luminosity does not trace the AGN mid-IR luminosity in the same fashion as seen in local Seyfert galaxies.

\section{Discussion and conclusion}\label{sec:con}

In this paper, we study the possible origins of the disparity between the different $L_X-L_{6\micron}$ relations found in the literature. We assemble samples of spectroscopically confirmed broad-line AGNs (type 1) across a wide range of X-ray survey areas
and depths to investigate the AGN intrinsic $L_X-L_{6\micron}$ relation
for luminous quasars. We test several observational constraints that could bias the observed $L_{\rm X}-L_{6\micron}$ relation, including intrinsic X-ray absorption, host galaxy contamination at mid-IR wavelengths, and X-ray survey flux limit. We argue that the most important factor that differentiates the \lxlsix\ relations from different studies is the X-ray survey flux limits (i.e., the Eddington bias), as we find that other factors do not affect the \lxlsix\ relation of luminous quasars significantly.

For the 2,509 X-ray detected AGNs in our sample,
we find that their \lxlsix\ relation could be well-described by the bilinear function 

\begin{equation}
\begin{aligned}
\log L_{6\micron}/{\rm erg\;s}^{-1} & < 44.79:\\
\log \frac{L_X}{{\rm erg\;s}^{-1}} =(0.84\pm0.03) &\log \frac{L_{6\micron}}{10^{45}\mathrm{erg\;s}^{-1}}+44.60\pm0.01\\
\log L_{6\micron}/{\rm erg\;s}^{-1} & \geq 44.79:\\
\log \frac{L_X}{{\rm erg\;s}^{-1}} = (0.40\pm0.03) & \log \frac{L_{6\micron}}{10^{45}\mathrm{erg\;s}^{-1}}+44.51\pm0.01,
\end{aligned}
\end{equation}
where the break luminosity is $\log L_{6\micron}/{\rm erg\;s}^{-1} \approx 44.79\pm0.11$.
For luminous quasars, the slope of their \lxlsix\ relation is significantly flatter than the
approximately linear relation observed in low- to moderate-luminosity AGNs \citep{gand09seyfir,asmu15},
which supports studies that suggest type 1 quasars have higher $L_{\rm MIR}$ to $L_{\rm X}$ ratios
than their local Seyfert counterparts \citep{fior09obsc,ster15}. The fit in Equation (2) does not take the X-ray non-detected AGNs into account, but we also show that the inclusion of X-ray non-detected AGNs does not affect the \lxlsix\ slope significantly given the high X-ray detection fractions of our samples. 

Since the rest-frame mid-IR emission in AGNs originates from hot dust heated by UV photons from the accretion disk, it is natural to consider the well-studied ratio between X-ray and UV monochromatic luminosities ($\alpha_{ox} = 0.38 (\log L_{2\mathrm{keV}}/L_{2500\mathrm{\AA\ }})$) to explain the luminosity-dependent \lxlsix\ relation. As pointed out by a number of studies \citep[e.g.,][]{tana79,stra05},
the \aox\ to AGN UV luminosity relation suggests that UV-luminous AGNs have relatively weak X-ray
emission compared to their less-luminous counterparts.
On the other hand, the radiation mechanism of AGN rest-frame mid-IR emission is driven by the UV photons from the accretion disk and the geometry of the dusty torus itself. 
While several dusty torus models
and observations have described the effect of increasing AGN UV luminosity on the geometry
of the dust distribution \citep{lawr91unified,luss13}, 
these models only predict a luminosity-dependent AGN obscured fraction and no 
drastic change of AGN UV to mid-IR spectral shapes is suggested. 
In fact, observational
studies by \cite{trei08} and \cite{luss10} have shown that the AGN mid-IR to
bolometric luminosity ratio is only weakly dependent on the AGN
bolometric luminosity. Studies of average SDSS quasar SEDs with different
infrared luminosities have also found marginal variation in the average
UV to mid-IR SEDs \citep{rich06,asse11qsolfunc}.

The relatively weak X-ray emission in luminous quasars has also been found in studies of X-ray bolometric corrections \citep[e.g.,][]{hopk07qlf,vasu07bolc,luss13}, which suggest that X-ray bolometric correction factors increases for AGNs with higher accretion efficiency (i.e., with higher Eddington rates).
The increase of X-ray bolometric correction factor in luminous quasars has also been implicitly suggested by the luminosity-dependent density evolution of the AGN X-ray luminosity
function \citep{aird10xlf}, as the rapid drop of the X-ray luminosity function implies that the most luminous X-ray AGNs are extremely rare. On the other hand, the AGN mid-IR luminosity function does not drop as significantly \citep[e.g.][]{brow06}.

If AGN X-ray and mid-IR luminosities followed the tight $L_{\rm X}-L_{6\micron} $ correlation seen in the local Seyfert galaxies, there would be as many as 40 type 1
quasars with $L_{X}>10^{44.7}$ erg s$^{-1}$ in the survey volume of \xbootes\ . However, according to the recent \cite{aird10xlf} X-ray luminosity function measured using deep X-ray surveys, the number of AGNs with $L_X>10^{44.7}$ erg s$^{-1}$ should only be 22.3 in the volume and redshift range of the Bo\"{o}tes dataset used in this work. In Bo\"{o}tes, there are only 18 sources more luminous than $L_X>10^{44.7}$ erg s$^{-1}$. Indeed, recent discoveries of the most infrared luminous AGNs in the universe, the hot dust-obscured galaxies \citep[hot-DOGs, e.g.][]{eise12hotdog,wu12wise,tsai15hotdog} have also found that their intrinsic X-ray luminosity is more consistent with the luminosity-dependent \lxlsix\ relations \citep[][]{ster14nustar,asse16,ricc16nustar}.

In conclusion, we have shown that for type 1 quasars, the relationship between AGN mid-IR and X-ray luminosities is not a simple power law, as has been observed for nearby Seyfert-luminosity AGNs. This result is crucial for the studies of AGN-galaxy coevolution, as the
dynamical range of X-ray luminosities is considerably smaller
than the dynamical range of AGN mid-IR luminosities.

\acknowledgments{Much of this work was in progress at the time of Steve Murray's untimely death in 2015. This work would not be possible without his generous GTO contribution to the \xbootes\ dataset and his strong support of \bootes\ multiwavelength surveys. He will be greatly missed by all of us.
We thank the referee for a careful reading and constructive suggestions that have improved the manuscript.
C.-T.J.C was supported by the William H. Neukom 1964 Institute for Computational Science.
R.C.H. acknowledges support from the National Science Foundation through AST award number 1515364 and from an Alfred P. Sloan Research Fellowship. R.J.A was supported by FONDECYT grant number 1151408. This publication makes use of data products from the Wide-field Infrared Survey Explorer, which is a joint project of the University of California, Los Angeles, and the Jet Propulsion Laboratory/California Institute of Technology, funded by the National Aeronautics and Space Administration. This publication also makes use of data products from the Two Micron All Sky Survey, which is a joint project of the University of Massachusetts and the Infrared Processing and Analysis Center/California Institute of Technology, funded by the National Aeronautics and Space Administration and the National Science Foundation. Funding for SDSS-III has been provided by the Alfred P. Sloan Foundation, the Participating Institutions, the National Science Foundation, and the U.S. Department of Energy Office of Science. The SDSS-III website is \url{http://www.sdss3.org/}.}

\bibliographystyle{apj}
\bibliography{../../BIB/chen14a_qsosf}{}

\end{document}